# Title: Finding the sources of missing heritability in a yeast cross

**Authors:** Joshua S. Bloom[1,2], Ian M. Ehrenreich[1,3], Wesley Loo[1,2], Thúy-Lan Võ Lite[1,2], Leonid Kruglyak[1,4,5*]

**Affiliations:**
[1] Lewis-Sigler Institute for Integrative Genomics, Princeton University, Princeton, NJ 08540

[2] Department of Molecular Biology, Princeton University, Princeton, NJ 08540

[3] Molecular and Computational Biology Section, University of Southern California, Los Angeles, CA 90089

[4] Department of Ecology and Evolutionary Biology, Princeton University, Princeton, NJ 08540

[5] Howard Hughes Medical Institute, Princeton University, Princeton, NJ 08540

*Correspondence should be addressed to L.K. (leonid@genomics.princeton.edu).

**Abstract**:

For many traits, including susceptibility to common diseases in humans, causal loci uncovered by genetic mapping studies explain only a minority of the heritable contribution to trait variation. Multiple explanations for this "missing heritability" have been proposed. Here we use a large cross between two yeast strains to accurately estimate different sources of heritable variation for 46 quantitative traits and to detect underlying loci with high statistical power. We find that the detected loci explain nearly the entire additive contribution to heritable variation for the traits studied. We also show that the contribution to heritability of gene-gene interactions varies among traits, from near zero to 50%. Detected two-locus interactions explain only a minority of this contribution. These results substantially advance our understanding of the missing heritability problem and have important implications for future studies of complex and quantitative traits.

**One Sentence Summary: In yeast, detected loci explain the additive contribution of genes to heritable variation in many quantitative traits.**

**Main Text:**

**Introduction**

Individuals within species show heritable variation for many traits of biological and medical interest. Most heritable traits follow complex inheritance patterns, with multiple underlying genetic factors (*1, 2*). Finding these factors has been a central focus of modern genetic research in humans, as well as in model organisms and agriculturally important species (*3-6*). Recent work, most notably genome-wide association studies (GWAS) in humans, has underscored the problem of "missing heritability"—although many genetic loci have been identified for a wide range of traits, these typically explain only a minority of the heritability of each trait, implying the existence of other, undiscovered genetic factors (*7*).

Multiple non-mutually-exclusive explanations have been proposed for missing heritability (*7*). One possibility is that the undiscovered factors could have effects that are too small to be detected with current sample sizes, or even too small to ever be individually detected with statistical significance (*8-10*). The existence of many small-effect variants is supported by studies showing that a large proportion of heritable trait variation is tagged when all GWAS markers are considered simultaneously (*11, 12*). Because GWAS can only detect variants that are common in the population, another possibility is that the undiscovered variants are too rare in the population to be captured by GWAS (*13*). One recent proposal suggests that non-additive interactions among loci (sometimes termed "epistasis") inflate heritability measures, creating "phantom" rather than missing heritability (*14*). Other proposed contributions to missing heritability include inflated heritability estimates, structural variation, gene-environment interactions, parent-of-origin effects, heritable epigenetic factors, and "entirely unforeseen sources" (*7, 15*). A better understanding of the sources of missing heritability is crucial for designing studies to find the missing components.

We set out to investigate these questions in the yeast *Saccharomyces cerevisiae*. We and others have previously used a cross between a lab strain and a wine strain to investigate the genetic basis of many complex traits (*16*), including global gene expression (*17*), protein abundance (*18*), telomere length (*19*), cell shape (*20*), gene expression noise (*21*), and drug sensitivity (*22*), and we have demonstrated "missing heritability" in this system (*23*). Recently, we used extreme quantitative trait locus (QTL) mapping (X-QTL) a bulk segregant approach that uses pools of millions of cross progeny (segregants), to detect many loci underlying heritable trait variation (*24*). We showed that for one trait, loci detected by X-QTL explained most of the heritability. However, pooled approaches do not allow direct estimates of heritability, the contribution of gene-gene interactions or locus effect sizes. Here we use a large panel of individually genotyped and phenotyped yeast segregants to accurately measure the heritable components of many quantitative traits, discover the underlying loci, and examine the sources of missing heritability.

**Heritability and the role of gene-gene interactions**

In order to estimate heritability and detect the underlying loci with high statistical power, we constructed a panel of 1008 prototrophic haploid segregants from a cross between a lab strain and a wine strain (Supplementary Fig. 1a,b and Supplementary Methods). These strains differ by 0.5% at the sequence level (*25*). We sequenced the parent strains to high coverage and compared the sequences to define 30,594 high-confidence single-nucleotide polymorphisms (SNPs) that distinguish the strains and densely cover the genome. We obtained comprehensive individual genotype information for each of 1008 segregants by highly multiplexed short-read sequencing (Supplementary Fig. 1c).

We sought to accurately measure a large number of quantitative traits in the segregant panel. To do so, we implemented a high-throughput end-point colony size assay and measured growth in multiple conditions, including different temperatures, pHs, and carbon sources, as well as addition of metal ions and small molecules (*16, 22, 24, 26*) (Supplementary Fig. 1d). We defined each trait as endpoint colony size normalized relative to growth on control medium, (Supplementary Methods) and obtained reproducible measurements with a strong heritable component for 46 traits (Supplementary Table 1). Most of the traits were only weakly correlated with each other (Supplementary Fig. 2).

Phenotypic variation in the segregant panel can be partitioned into the contribution of heritable genetic factors (broad-sense heritability) and measurement errors or other random environmental effects. Broad-sense heritability can in turn be partitioned into the contribution of additive genetic factors (narrow-sense heritability), dominance effects, gene-gene interactions, and gene-environment interactions (*27, 28*). In our experiment, dominance effects are absent because the segregants are haploid, and gene-environment interactions for a given trait should also be absent as all the segregants are grown simultaneously under uniform conditions. Thus our estimates of broad-sense heritability include additive and gene-gene interaction components, while our estimates of narrow-sense heritability include only the additive component. The difference between the two heritability measures (recently termed "phantom heritability" (*14*)) therefore provides an estimate of the contribution of gene-gene interactions.

We estimated broad-sense heritability from repeatability of trait measurements (Supplementary Methods). Estimating narrow-sense heritability usually involves measuring phenotypic similarity for different degrees of relatedness. We took advantage of a recently developed genomic approach in which narrow-sense heritability is estimated by comparing phenotypic similarity among individuals with their actual genetic relatedness, computed from dense genotype data (Supplementary Methods) (*29*). Among the 46 traits, broad-sense heritability estimates ranged from 0.40 to 0.96, with a median of 0.77. Narrow-sense heritability estimates ranged from 0.27 to 0.85, with a median of 0.55 (Fig. 1). We used the difference between these measures to estimate the fraction of genetic variance due to gene-gene interactions, which ranged from 0.02 to 0.50 with a median of 0.25. Thus, the genetic basis for variation in some traits is almost entirely due to additive effects, while for others up to half of the heritable component is due to gene-gene interactions.

**Additive heritability explained by detected loci**

Next, we sought to map the additive heritable variation to specific QTL. Simple linkage analysis of one marker at a time revealed multiple QTL per trait (Supplementary

Methods). To more accurately capture the effects of each QTL while controlling for the other QTL affecting the same trait, we used a step-wise forward-search approach to detect QTL and build a multiple-regression model (Supplementary Methods). With this approach, we detected a total of 591 QTL for 46 traits at an empirical false-discovery rate (FDR) of 5%. We observed varying degrees of trait complexity, with a minimum of 5, a maximum of 29, and a median of 12 QTL per trait. These numbers of QTL are comparable to those previously seen for a smaller set of traits by X-QTL (*24*). Consistent with theoretical predictions (*30*) and previous observations, we detected many more QTL of small effect than of large effect (Fig. 2). Some traits showed a distribution of QTL effect sizes roughly consistent with Orr's evolutionary model (*30*), whereas others showed one or more larger-than-expected QTL (Supplementary Fig. 3).

Having identified QTL, we next measured the fraction of additive heritability explained by our model of detected QTL for each trait. To obtain unbiased estimates, we performed 10-fold cross-validation by detecting QTL in a subset of the segregant panel and estimating the effects in the rest of the panel. Across the traits, the detected loci explained between 72% and 100% of the narrow-sense heritability, with a median of 88% (Fig. 3a). The detected loci explained between 41% and 84% of the broad-sense heritability, with a median of 67%, and between 26% and 80% of the total phenotypic variance, with a median of 49%. Thus, high statistical power provided by the large segregant panel allowed us to detect QTL that jointly explain most of the additive heritability for the traits studied here. By analyzing subsets of the data, we showed that "missing" narrow-sense heritability can be explained by insufficient sample sizes (Fig.

3b). For instance, we detected 16 significant QTL, which jointly explain 78% of narrow-sense heritability for growth in E6-berbamine, in a panel 1000 segregants, (Fig. 4a), but only 2 of these, explaining 21% of narrow-sense heritability, also reached statistical significance in a smaller panel of 100 segregants (Fig. 4b). For traits with mostly additive genetics, the high fraction of variance explained by the detected QTL allowed us to accurately predict individual trait values from QTL genotypes (Fig. 5).

**Detecting two-locus interactions**

Differences between the estimates of broad-sense and narrow-sense heritability for many traits imply the presence of genetic interactions. We next sought to identify specific two-locus interactions. For each trait, we first performed an exhaustive two-dimensional (2D) scan for pairwise interactions. At a LOD score of 6.2, corresponding to an empirical FDR of 10%, we detected significant QTL-QTL interactions for 17 of the 46 traits, with a total of 23 interacting locus pairs. A 2D scan has low statistical power due to the large search space. Power can be increased, at the cost of missing interactions between loci with no main effects, by testing only for interactions between each locus with significant additive effects and the rest of the genome (*31*). Using this approach, we detected interactions for 24 of the 46 traits, with a total of 78 QTL-QTL interactions at an FDR of 10%. We observed a minimum of 1 and a maximum of 16 pairwise interactions per trait. These 78 pairs included 20 of the 23 locus pairs detected in the exhaustive 2D scan, suggesting that two-locus interactions in which neither locus has a detectable main effect are uncommon. For 47 of the 78 pairs, both loci were detected as significant in the single-locus search for additive effects. In the remaining 31 cases, the additive effect of the second locus was too small to reach genome-wide significance, although it was nominally significant in 10 of these cases. These observations are broadly consistent with our previous work on genetic interactions that affect gene expression traits (*31, 32*).

For most of the traits with a sizeable difference between broad-sense and narrow-sense heritability, pairwise interactions were either not detected or explained little of the difference (Fig. 6). The detected interaction effects were typically small (a median of 1.1% of genetic variance per interaction or a median of 3% of genetic variance per trait). Only in a few cases did detected genetic interactions explain much of the difference between broad-sense and narrow-sense heritability. Most notably, in the case of growth on maltose, one strong interaction explained 14% of the genetic variance and 76% of the difference between broad-sense and narrow-sense heritability (Fig. 6 inset).

**Discussion**

We have used a large panel of segregants from a cross between two yeast strains to investigate the genetic architecture of 46 quantitative traits. We measured both the total and the additive contributions of genetic factors to trait variation, and showed that these often differ. The observed differences between total and additive heritability estimates suggest that the contribution of genetic interactions to broad-sense heritability ranges from zero to 50%. However, with a few exceptions, the specific combinations of loci that account for these interactions remain elusive. There are several possible explanations for this result. First, the statistical power to detect interactions is lower than the power to detect main effects. Second, individual interaction effects are expected to be smaller than additive effects, and hence their detection requires even larger sample sizes (*14*). Finally, higher-order interactions among more than two loci could also contribute (*33*). Our estimates of the contribution of interactions in a cross may overestimate their contribution to trait heritability in a population, because a higher proportion of variance is expected to be additive as allele frequencies depart from one-half (*34*).

The large size of the panel allowed us to detect specific loci that jointly account for the great majority of the additive (narrow-sense) heritability of each trait (72-100%). Human traits examined by GWAS vary in their genetic complexity (*7*), ranging from macular degeneration, for which 5 variants in 3 genes explain roughly half of the genetic risk (*35*), to height, for which 180 loci explain about 13% of heritability (*36*). However, compared to our results in a yeast cross, GWAS typically detect a larger number of loci explaining a smaller proportion of trait heritability. One obvious difference is that the number of variants segregating in a cross between two strains is smaller than the number of common variants segregating in a population sample. The difference is roughly a factor of three for a neutral allele frequency spectrum, and potentially much larger if functional variants are deleterious and hence shifted toward lower frequency (Supplementary Methods). The human genome also offers a larger target size, perhaps by a factor of five, for variants affecting a trait (Supplementary Methods). These very rough estimates suggest that we might expect at least 15 times more loci to be found by GWAS than the median of 12 loci per trait we observe in the yeast cross. Because of the resolution of linkage analysis in our cross, some QTL may contain multiple linked variants, further increasing the true number of loci. Several additional factors could lead to a larger "missing heritability" in humans: the fraction of heritability due to genetic interactions could be higher (*14*), rare variants may account for a disproportionately large contribution of heritable variation (*37-39*), and some human traits might be inherently more complex than yeast traits in that they integrate over physiological processes involving a larger number of underlying gene pathways. Within-locus dominance effects represent an additional source of genetic complexity in diploid organisms.

Our results are consistent with the suggestions that missing additive (narrow-sense) heritability arises primarily from many loci with small but not infinitesimal effects. These loci can be discovered in studies with sufficiently large sample sizes, although the optimal study designs will depend on the population frequency spectra of the causative alleles. Because all alleles are fixed at a frequency of one-half in a cross, we cannot yet delineate the contributions of common and rare variants to inherited variation, but we plan to do so in future studies.

**Supplementary Materials**

Materials and Methods

Figs. S1, S2, S3

References (40-54)

Table S1, S2

**Acknowledgements:**

We thank David Botstein, Megan McClean, Erik Andersen, Frank Albert, Sebastian Treusch, Rajarashi Ghosh, and Xin Wang for comments on the manuscript, Yue Jia for technical assistance, and Eric Lander for discussions. This work was supported by NIH grant R37 MH59520, a James S. McDonnell Centennial Fellowship, and the Howard Hughes Medical Institute (L.K.), an NSF fellowship (J.S.B.), NIH postdoctoral fellowship F32 HG51762 (I.M.E.), and NIH grant P50 GM071508 to the Center for Quantitative Biology at the Lewis-Sigler Institute of Princeton University.

**Figure 1**

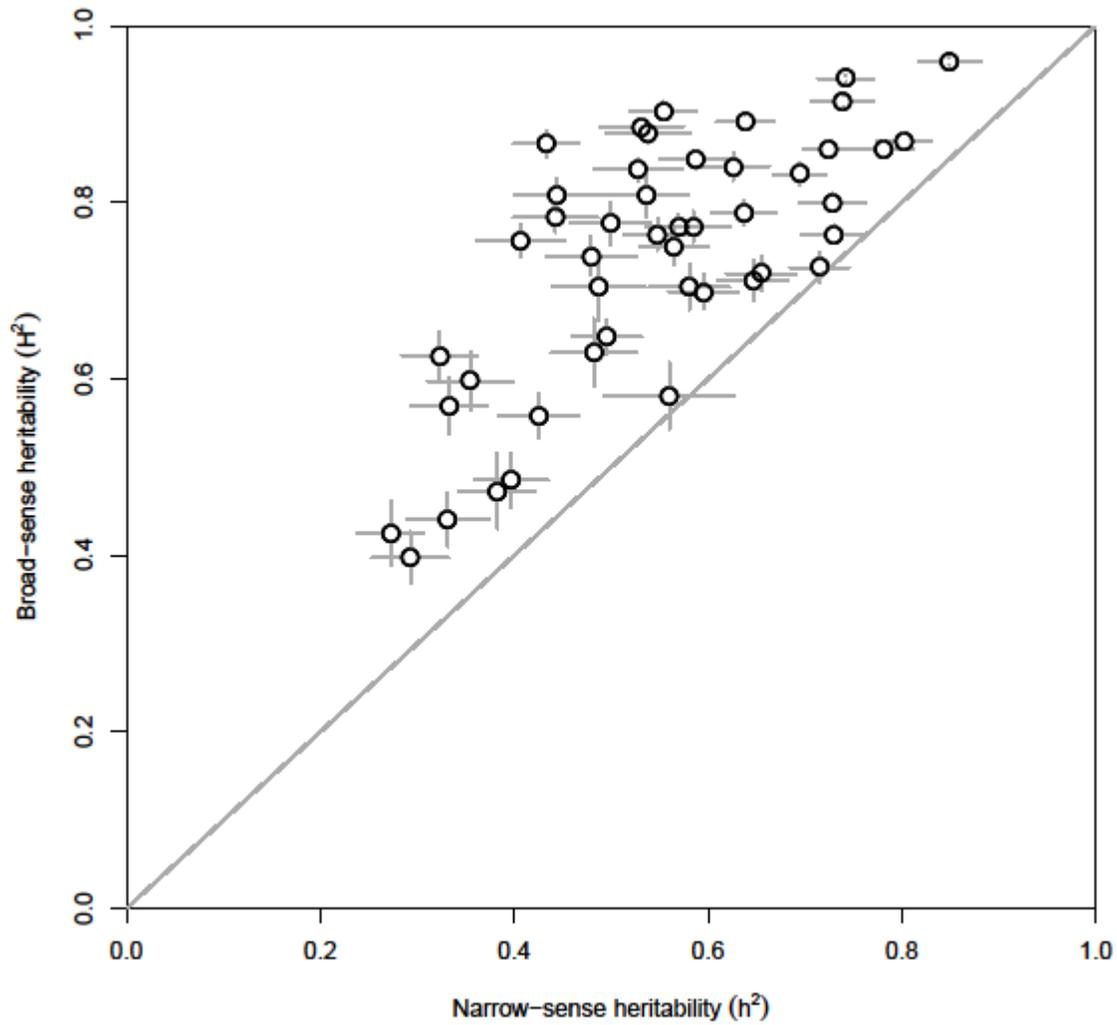

**Fig.1 | Heritability for 46 yeast traits.**

The broad-sense heritability ($H^2$) for each trait (Y-axis) is plotted against the narrow-sense heritability ($h^2$; X-axis). Error bars show standard errors in heritability estimates. The diagonal line represents $H^2 = h^2$ and is shown as a visual guide.

**Figure 2**

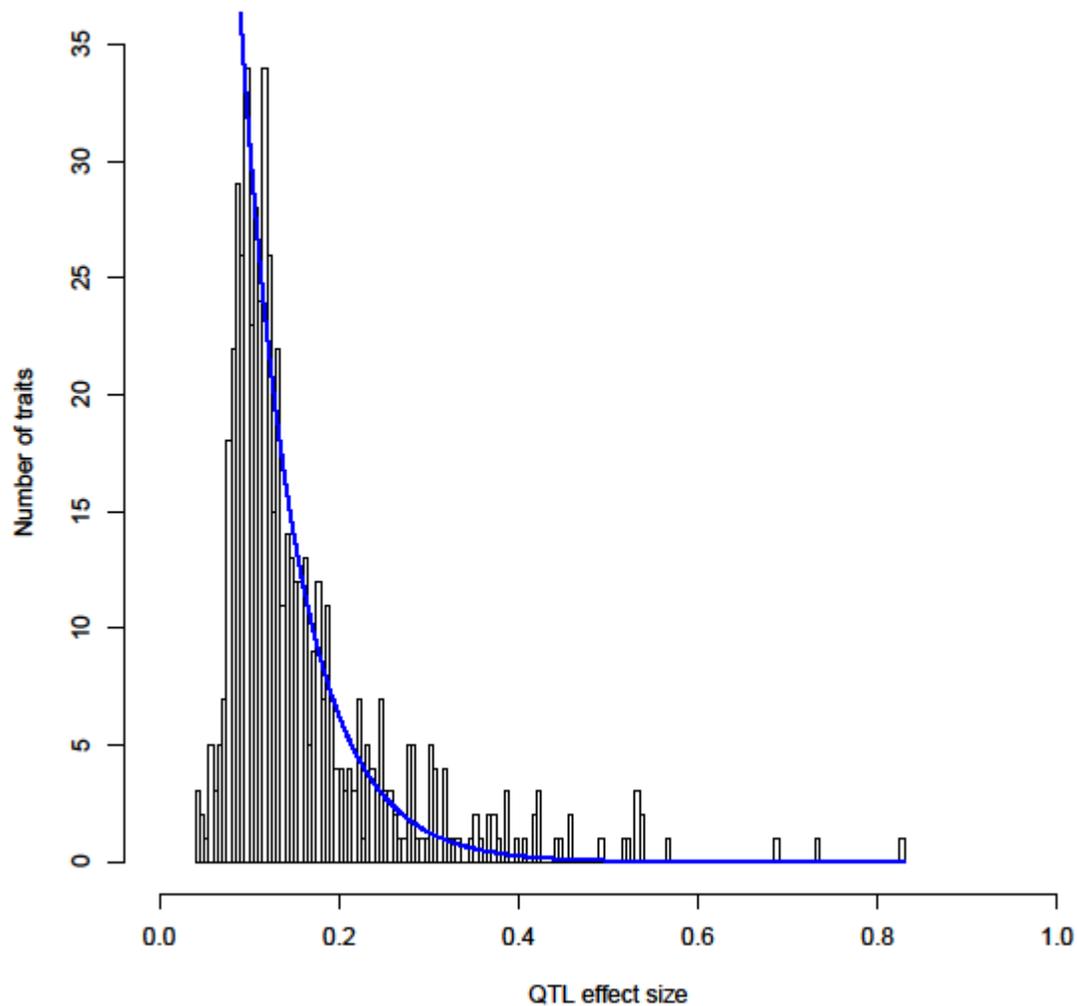

**Fig.2 | Distribution of QTL effect sizes.**

A histogram of QTL effect sizes across all traits is plotted, showing that most detected QTL have small effects. Effect size here is the absolute value of the standardized difference in allelic means for each QTL. The blue line indicates a fit of a truncated exponential distribution of effect sizes.

**Figure 3**

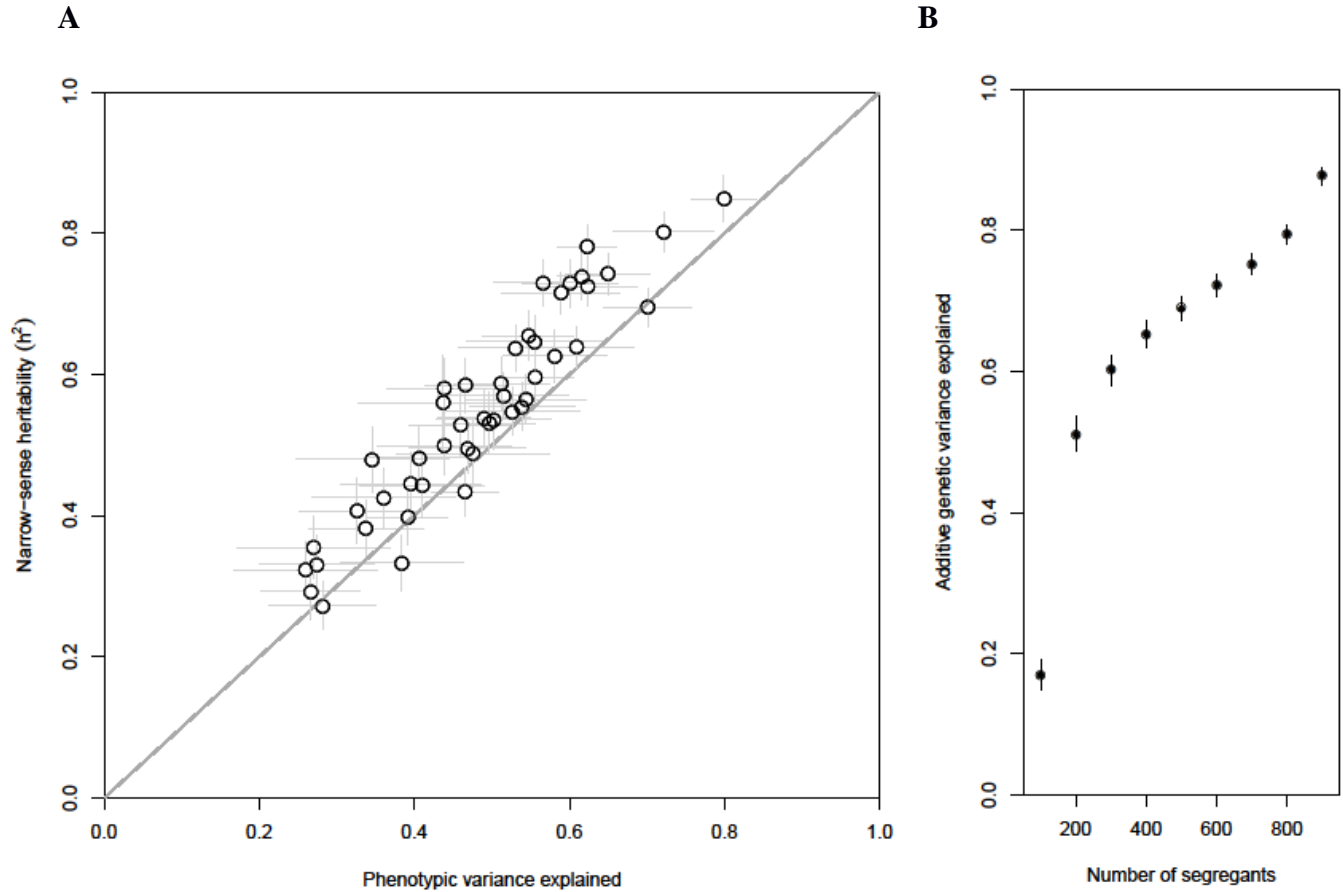

**Fig.3 | Most additive heritability is explained by detected QTL.**

(A) The narrow-sense heritability ($h^2$) for each trait (Y-axis) is plotted against the total variance explained by detected QTL (X-axis). Error bars show standard errors. The diagonal line represents $h^2$ = (variance explained by detected QTL) and is shown as a visual guide. (B) The average fraction of additive genetic variance explained (Y-axis) is plotted against number of segregants used for QTL detection (X-axis). Error bars show standard errors.

**Figure 4**

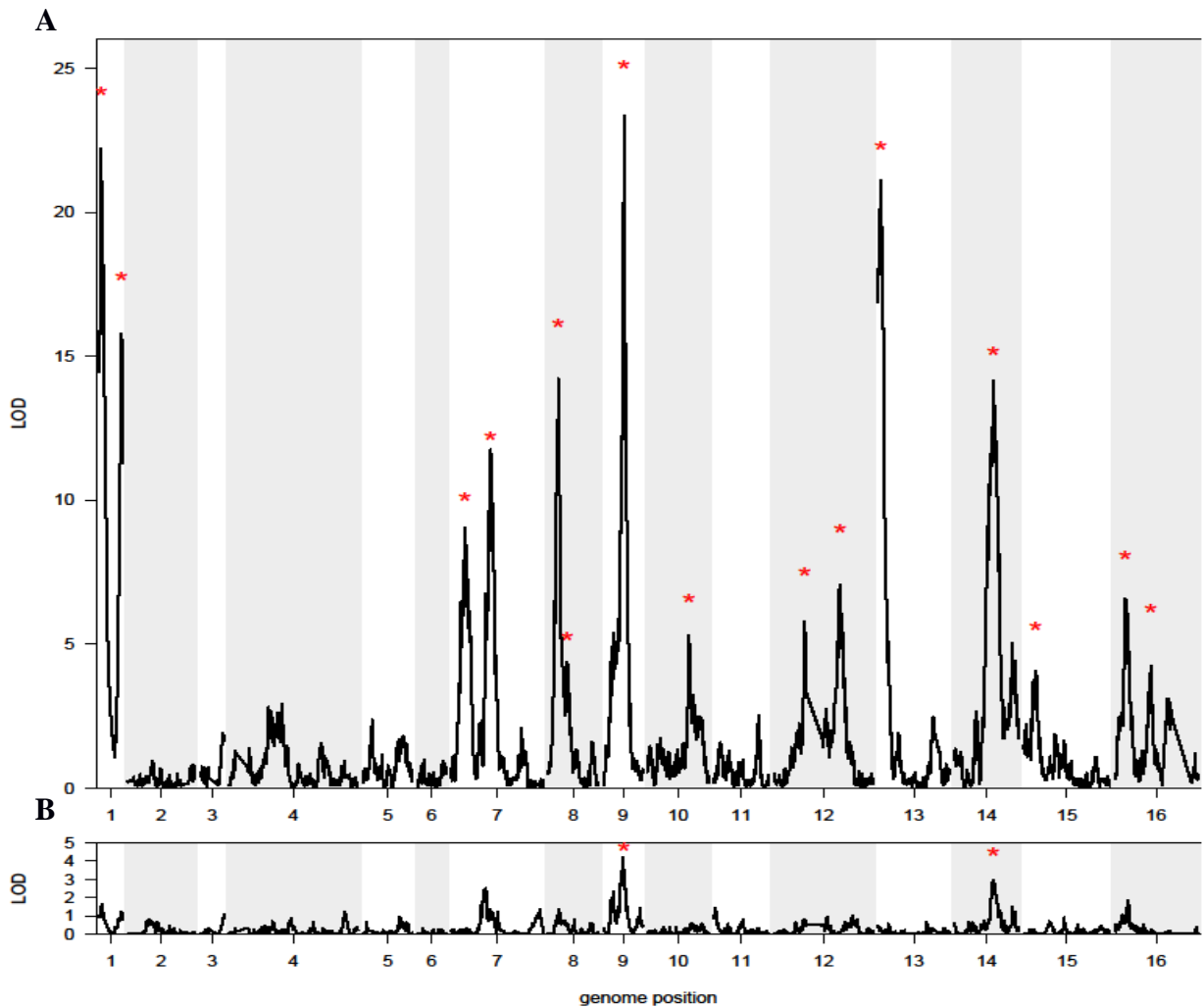

**Figure 4 | QTL detection for a complex trait.**

LOD score ($\log_{10}$ of the odds ratio for linkage) is plotted against the genetic map. Red stars indicate statistically significant QTL. (A) LOD score plot with 1005 segregants for growth in E6-berbamine. (B) LOD score plot with 100 segregants for growth in E6-berbamine. The 15 significant QTL in A explain 78% of the narrow-sense heritability, compared with 21% for the 2 significant QTL in B.

**Figure 5**

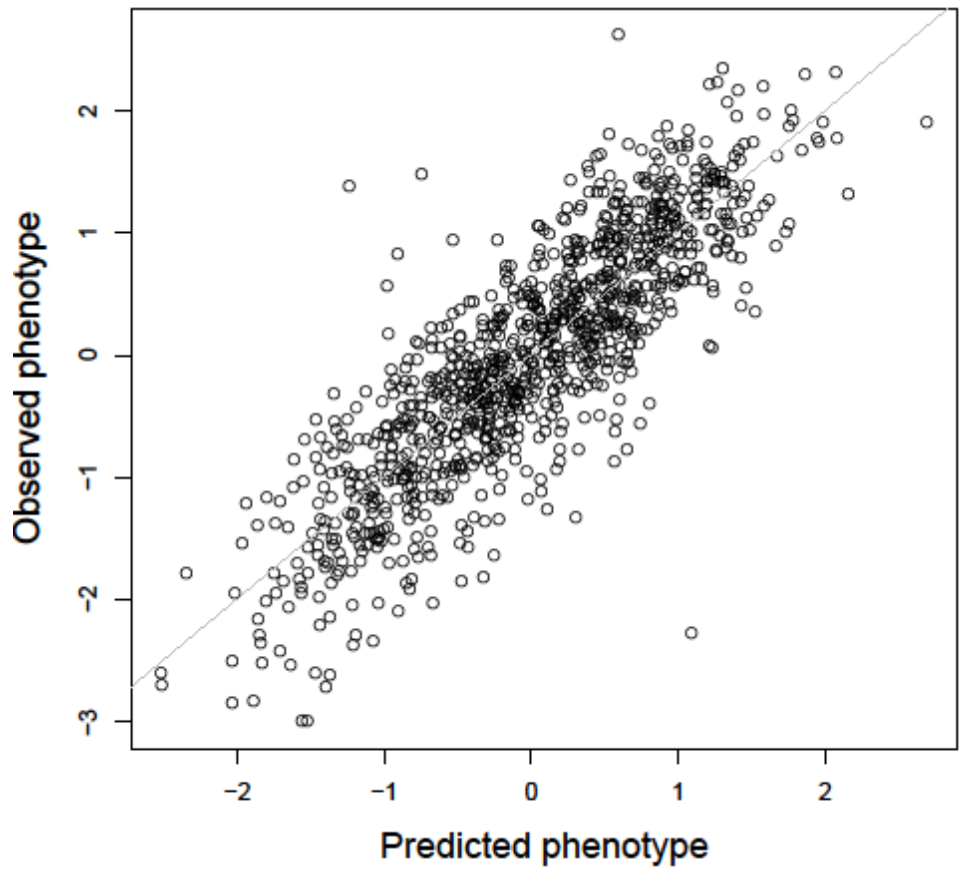

**Figure 5 | Prediction of segregant trait values from QTL phenotypes.**

The observed phenotypic values for growth in lithium chloride (Y-axis) are plotted against the predicted phenotypic values based on a cross-validated additive model of 22 QTL. The additive QTL model explains 90% of the narrow-sense heritability and 83% of the broad-sense heritability. The diagonal line represents (observed phenotype) = (predicted phenotype) and is shown as a visual guide.

**Figure 6**

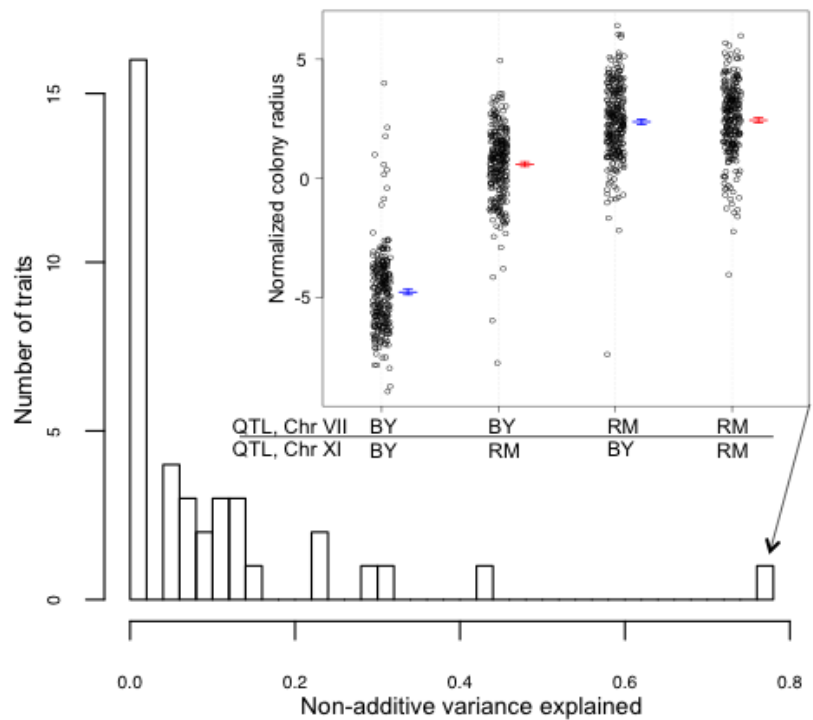

**Figure 6 | Non-additive variance explained by QTL-QTL interactions.**

A histogram of the fraction of non-additive variance explained by detected QTL-QTL interactions per trait is plotted. (Inset) Phenotypes for growth in maltose are shown, grouped by two-locus genotypes at the two interacting QTL on chromosomes VII and XI. This QTL-QTL interaction explained 76% of the difference between broad-sense and narrow-sense heritability.

**Supplementary Materials:**

**Materials and methods:**

**Construction of segregant panel**

We crossed a prototrophic BY parent (derived from a cross of BY4716 and BY4700) that is MAT**a** with a prototrophic RM parent (derived from RM11-1a) that is MATα *ho∆::HphMX4 flo8∆::NatMX4 AMN1-BY*. Diploid zygotes were recovered and sporulated for 5-7 days in SPO++ sporulation medium (http://dunham.gs.washington.edu/sporulationdissection.htm) in a roller drum at room temperature. Tetrads were dissected (*40*) using the MSM 400 dissection microscope (Singer Instrument Company Ltd.). Colonies from four-spore tetrads were innoculated into 150 μL of YNB + 2% glucose in 96 well plates (Corning; 07-200-656), grown for 48 hours at 30° without shaking, and stored as frozen stocks in 20% glycerol. 1056 four-spore tetrads that displayed 2:2 segregation of hygromycin resistance, G418 resistance, and mating type were retained. A Biomek FX (Beckman Coulter) was used to select one segregant from each four-spore tetrad for downstream analysis.

**Power calculations**

We calculated statistical power (1- β) for sample sizes of 100 and 1000 segregants in R using the 'power.t.test' function (*41*). Power was calculated over a range of effect sizes, where effect size was calculated as the percent genetic variance explained by a single QTL. To correct for multiple testing over thousands of markers across the genome, a genome-wide significance threshold (α) of $p < 2*10^{-4}$ was used.

**DNA preparation and sequencing library construction**

Segregants were innoculated into 1 mL deep-well 96-well plates (Thermo Scientific; 12-565-395) in 800 μL of YPD and grown 2 days at 30° without shaking. Plates were sealed with Breathe-Easy gas-permeable membranes (Sigma-Aldrich; Z380059). DNA was extracted using 96-well DNEasy Blood & Tissue kits (Qiagen; 69582). DNA concentrations were determined using the Quant-it dsDNA high sensitivity DNA quantification kit (Invitrogen; Q-33120) and the Biotek Synergy 2 plate reader. DNA was diluted to 1.66 ng/μL. Per sample, 15 μL of 1.66 ng/μL DNA was added to 4 μl of 5X Nextera HMW buffer, 0.95 μl of water, and 0.05 μl of Nextera Enzyme Mix (Epicentre; GA0911-96). The transposition reaction was performed for 5 minutes at 55°. 100 μl of water was added to each sample, and the samples were purified using the MinElute kit (Qiagen; 28053). 15 μl of the purified fragmented DNA was PCR-amplified and barcoded with custom 5 bp sequences using ExTaq polymerase (Takara, RR001A) and between 15 and 30 PCR cycles. 5 μl of each PCR-amplified sample was combined into one 96-plex library. The combined library was loaded on a 2% agarose gel, and the 350 bp to 650 bp region was excised and gel-extracted using QiaQuick Gel Extraction Kit (Qiagen; 28704). Final libraries were diluted to 3.3 ng/μl and sequenced using the single-end module on a HiSeq 2000 (Illumina) with 100 bp reads.

**Determining segregant genotypes**

Custom R and python code was used to demultiplex the sequencing data and trim ends. Sequencing reads were assigned to segregants based on the 5 base-pair (bp) barcode at the beginning of each read. The internal 19 bp transposon sequence and 10 bp on the right end of each read were removed. Reads were aligned to the S288C reference genome using the Burrows-Wheeler Aligner (BWA) (*42*) with the '-q 30' parameter. SAMTools(*43*) was run with the 'view' command and '-bHsq 1' parameters to retain uniquely mapping reads. Sequence variants were identified using SAMTools (*43*) with the 'mpileup' command and parameters '-d 10000 -D -u'. 42,689 high-confidence sequence variants between BY and RM were determined from sequencing the parental strains at greater than 50-fold coverage. Variants in the segregants were restricted to these 42,689 expected sites using 'bcftools view' (*43*) with the parameters '-N -c -g -v -P flat'. Genotype likelihoods for the BY and RM alleles for each genotypic variant were extracted from the VCF file using custom R code.

For each segregant and chromosome, a hidden Markov model (HMM) was used to calculate the posterior likelihood that the read data was coming from the BY allele, the RM allele, a BY gene-conversion event or an RM gene-conversion event. Genotypes were called as the BY variant if the $\log_{10}$ ratio of the BY posterior likelihood and the RM posterior likelihood was greater than 2, the RM variant if this ratio was less than -2, and missing data if between -2 and 2. 1,008 out of 1,056 segregants had between 25 and 120 recombination breakpoints and at least 35,000 markers with genotype calls, and these were retained for downstream analysis. Genotypic markers were excluded if their allele frequency was greater than 56% or less than 45%, or if they were not called in 99% of the segregants. This resulted in a final set of 30,594 genotypic markers. Markers with missing data were imputed using the Viterbi algorithm as implemented in the R/qtl package (*44*). Adjacent markers with the same genotypes in all segregants were collapsed to one unique marker, resulting in a final set of 11,623 unique genotypic markers.

**Phenotyping by end-point growth on agar plates**

Individual segregants were inoculated in at least 2 different plate configurations into 384-well plates (Themo Scientific; 264574) with 50 mL of YPD and grown for 36 to 48 hours in a 30° degree incubator without shaking. Each target agar plate was made with 50 mL of media (YPD or YNB) and with drug or condition of choice (Supplementary Table 1). The Singer Rotor HDA pinning robot was used to pin the segregants to the agar plates. Before pinning, each 384-well plate was mixed for 1 minute at 2000 RPM using a MixMate (Eppendorf). Segregants were pinned to the agar plates with 100% pin pressure and 384 long pins (Singer Instruments; RP-MP-3L). After pinning, the plates were incubated at 30°, or the specified condition temperature (Supplementary Table 1), for approximately 48 hours. Plates were scanned face-up and without lids on an Epson 700 transparency scanner with 400 dpi resolution and a greyscale bit-depth of 8. Images were saved as TIFFs or 99% quality JPEGs. The pixel coordinates of the centers of the four corner colonies for each plate were manually identified using ImageJ (*45*).

Custom R code was written to determine the size of each colony. Expected colony positions were calculated using manually identified coordinates of the corner colonies for each plate. Images were segmented using k-means clustering on the distribution of pixel intensities across a plate. The Voronoi region of each 9-pixel diameter circular seed, corresponding to the expected location of a colony, was used together with the segmented image to match colonies to their expected positions. This was implemented using functions in the EBImage (*46*) R package. The radius of each colony was calculated as $\sqrt{\frac{pixel\_count}{\pi}}$. Colonies with more than 15 pixels touching the edge of the image and colonies larger than 3,500 pixels were removed and treated as missing data in downstream analysis. Irregular colonies, representing image processing or pinning artifacts, were removed if $\frac{perimeter}{2 * \sqrt{pixel\_count}} > 1$ and radius>20. These irregular colonies were also treated as missing data for downstream analysis. Colonies on each edge of a plate were tested for difference with all other colonies on the plate using a Wilcoxon rank-sum test. All colonies on an edge with p<0.05 were treated as missing data and excluded from downstream analysis. Images were further inspected manually, and colonies subject to pinning or image-processing artifacts were removed. To normalize for occasional subtle within-plate spatial growth artifacts, a robust locally weighted regression was fit to the radius measurements using functions in the locfit R package (*47*). The residuals were used for downstream analysis. End-point growth measurements were normalized for growth on control media by fitting a regression for effect of growth on control media and using the residuals for downstream analysis. 140 conditions were assayed. If multiple doses of a compound were tested, the dose with the highest heritability and phenotype data for at least 600 segregants was retained. Traits with narrow-sense heritability less than 25% were excluded from downstream analysis.

**Calculating heritability**

Broad-sense heritability was calculated using replicated segregant data and a random effects ANOVA. This was implemented using the 'lmer' function in the lme4 R package (*48*). The variance components $\sigma_G^2$, the genetic variance due to effect of segregant, and $\sigma_E^2$, the error variance, were calculated, and broad sense-heritability was estimated as $\frac{\sigma_G^2}{\sigma_G^2 + \sigma_E^2}$. Standard errors were calculated by delete-one jackknife.

Narrow-sense heritability was calculated for each trait using a linear mixed model (*49*). This can be written as $\mathbf{y}=\beta\mathbf{1_N}+\mathbf{Zu}+\mathbf{e}$. Here $\mathbf{y}$ is a vector of phenotype values for N segregants, $\beta$ is the overall mean, and $\mathbf{1_N}$ is a vector of N ones. Z is an NxN identity matrix, $\mathbf{u}$ is a vector of random effects (BLUPs or breeding values for each segregant), and $\mathbf{e}$ is a vector of residuals. The variance structure of the phenotypes is written as $\mathbf{V}=\mathbf{A}\sigma_A^2 + \mathbf{I}\sigma_{EV}^2$, where $\mathbf{A}$ is the relatedness matrix between all pairs of segregants, estimated from our genotype data as the proportion of markers shared IBD between all pair of segregants, $\mathbf{I}$ is an NxN identity matrix, $\sigma_A^2$ is the polygenic additive genetic variance explained by the SNPs, and $\sigma_{EV}^2$ is the error variance. Variance components were estimated using the rrBLUP R package (*50*), and narrow-sense heritability was estimated as $\frac{\sigma_A^2}{\sigma_A^2 + \sigma_{EV}^2}$. Standard errors were calculated by delete-one jackknife.

Although the average genetic relatedness among the segregants is 0.5, it varies due to random Mendelian segregation, and the actual relatedness for any pair of segregants can 25be calculated from high-density genotype data as the proportion of SNP alleles shared by these segregants. In our segregant panel, the standard deviation of relatedness was 0.048.
**Mapping additive QTL**

46 traits were chosen for QTL mapping based on the criteria described above. Each trait was scaled to have mean 0 and variance 1. We tested for linkage by calculating LOD scores for each genotypic marker and each trait as $-n(\ln(1 - R^2)/2\ln(10))$, where R is the Pearson correlation coefficient between the segregant genotypes at the marker and segregant trait values (*28*).

To estimate significance empirically, assignment of phenotype to each segregant was randomly permuted 1000 times while maintaining the correlation structure among phenotypes. The maximum LOD score for each chromosome and trait was retained (*51*). The FDR was calculated as the ratio of expected peaks to observed peaks across different LOD thresholds. Genetic markers corresponding to QTL peaks which were significant at an FDR of 5% were added to a linear model for each trait. Trait-specific linear models that included the significant QTL genotypes as additive covariates were computed, and phenotypic residuals were estimated. Phenotypic residuals for each trait were then used for another round of QTL detection (*52*). This process of peak detection, calculation of empirical significance thresholds, and expansion of the linear model for each trait to include significant QTLs detected at each step was repeated 4 times. The LOD thresholds corresponding to a 5% FDR at each step were 2.68, 2.92, 3.72 and 4.9.

**Calculating effect size and variance explained by additive QTL**

For each trait, a multiple regression linear model was computed with trait-specific QTL genotypes as independent variables. Phenotypes for each trait were scaled to have mean 0 and variance 1. The multiple regression coefficients are the standardized differences in allelic means for each QTL while controlling for the effects of other segregating QTL. The square of the multiple regression coefficient is the fraction of phenotypic variance explained by a QTL. The fitted truncated exponential distribution in Figure 2 is parameterized as $\frac{bnw}{e^{-lb} - e^{-rb}} e^{-bx}$, where x is the absolute value of the multiple regression coefficient pooled across traits and QTL, n is the number of bins, w is the bin size, l and r are left and right truncation points and b is estimated using maximum likelihood. Here l=0.12 and r=0.35 to correspond to the magnitude of effect sizes where power is nearly 100% and to exclude large-effect QTL.

The total phenotypic variance explained by the multiple regression QTL model is the $R^2$ from the model, which was calculated using the 'fitqtl' function in R/qtl. Unbiased estimates of the total phenotypic variance explained by the multiple QTL model were calculated by standard 10-fold cross validation. The segregants were randomly split into 10 equal-sized groups, 9 groups were combined for QTL detection using the algorithm described above, and the remaining group was used to estimate QTL effect sizes. This process was repeated for each of the 10 groups, and the average of the 10 estimates was calculated.

**Detecting QTL-QTL interactions**

For computational efficiency, the marker set was reduced to 4,420 by picking one physical marker closest to each centimorgan (cm) position on the genetic map. For the full genome scan for interacting QTL, a LOD score corresponding to the likelihood ratio of a model that includes an interaction term, **y**=a**x**+b**z**+c**xz**+d, to a model that does not, y=a**x**+b**z**+d, was computed for each trait and every marker pair. Here, **y** is the residuals vector for each trait after fitting the additive QTL model, **x** is the genotype vector at one position in the genome, **z** is the genotype vector at another position in the genome at least 25 cm away from **x**, and a,b,c, and d are estimated parameters specific to each trait and marker pair (*31, 32*). FDR at different LOD thresholds was calculated by dividing the average number of peaks obtained in 1000 permutations of the data that scramble the segregant identities by the number of peaks observed in the real data.

To increase statistical power, we tested for interactions between each locus with significant additive effects and the rest of the genome. A LOD score for interaction was computed in exactly the same manner as for the full 2D scan described above, except that **x** was constrained to genotypes corresponding to trait-specific significant additive QTL. FDR was calculated as above.

**Conversion factors for yeast vs. human expected numbers of QTL**

The number of variants segregating in a cross between two strains is smaller than the number of (common) variants segregating in a population sample. If the causal variants follow a neutral allele frequency spectrum, then under the standard neutral model of population genetics, the relationship between the number of variants segregating in a cross of two haploid strains ($S_2$) and the number of common variants segregating in a population ($S_f$) with minor allele frequency f is $S_f = S_2 \ln(\frac{1-f}{f})$ (ref. (*53*)). By setting f to 0.05, we obtain a ratio $S_f/S_2$ of approximately three (2.94). The ratio is 2.2 for f = 0.1 and 4.6 for f = 0.01. If the functional variants are at least weakly deleterious, and hence skewed toward lower frequency, the conversion factors will be larger.

Several approaches can be used to obtain a rough idea of relative target size for mutation. The yeast genome is 12 Mb, whereas the coding regions of the human genome alone are ~30 Mb, and at least the same amount of non-coding sequence is expected to be functional based on selective constraint (*54*). If we assume that the number of loci involved in a trait is proportional to the number of bases available for mutations with functional consequences, human traits should be more complex than yeast traits by roughly a factor of 5. Similarly, yeast has ~5700 genes compared to ~20,000 for humans, which gives a conversion factor of ~3.5.

**Supplementary Figures:**

**Figure S1.**

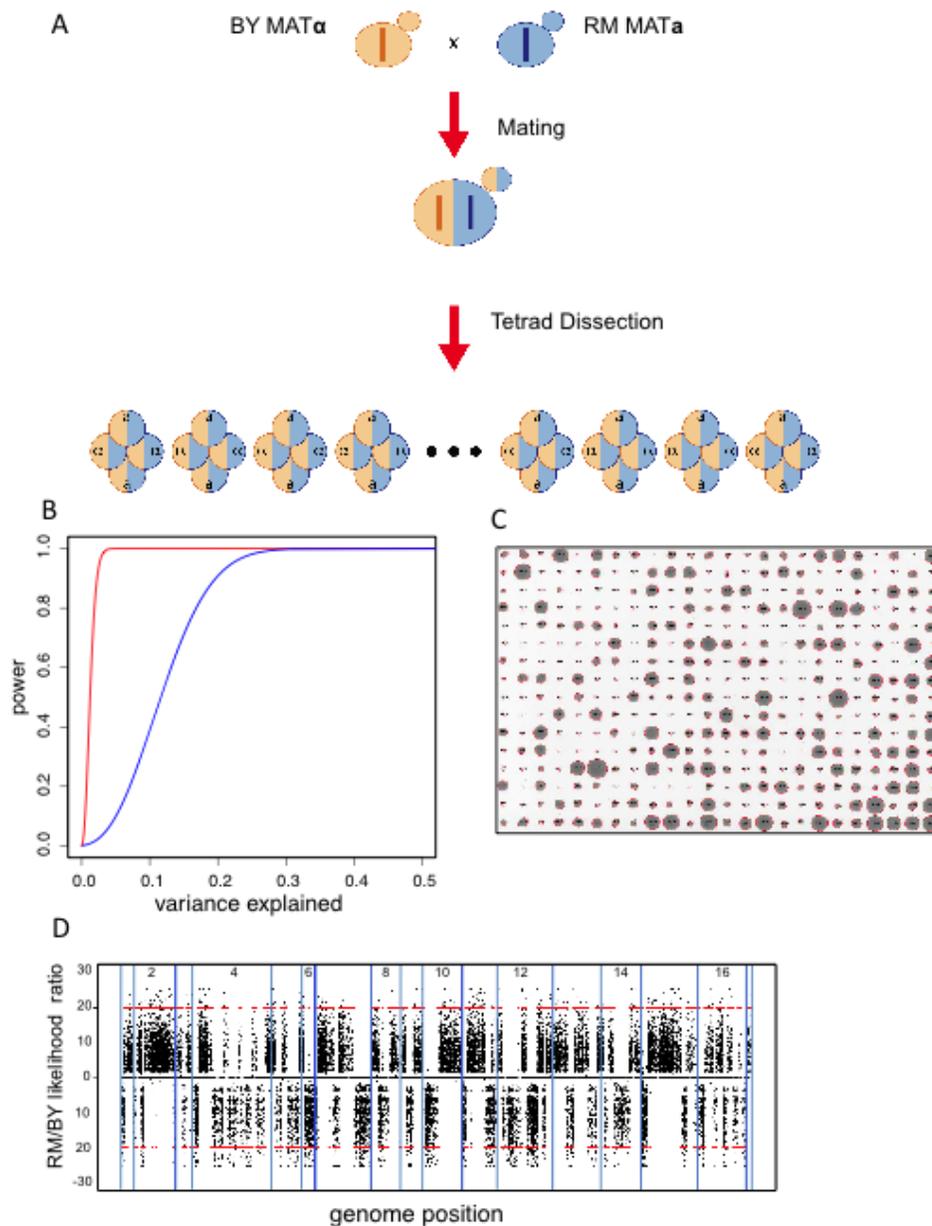

The design of the segregant panel is shown in (A). (B) Curves illustrating statistical power are shown for mapping populations of 100 (blue) and 1000 (red) segregants at a genome-wide significance threshold. (C) An image of endpoint colony growth is shown for 384 segregants, with the outlines of colonies, as detected by our image processing software, indicated in red. (D) Parental genotype likelihood ratios are plotted (Y-axis) against genome position (X-axis) for a representative segregant; the red bars indicate parental haplotype calls, and the vertical blue bars delineate chromosomes.

**Figure S2.**

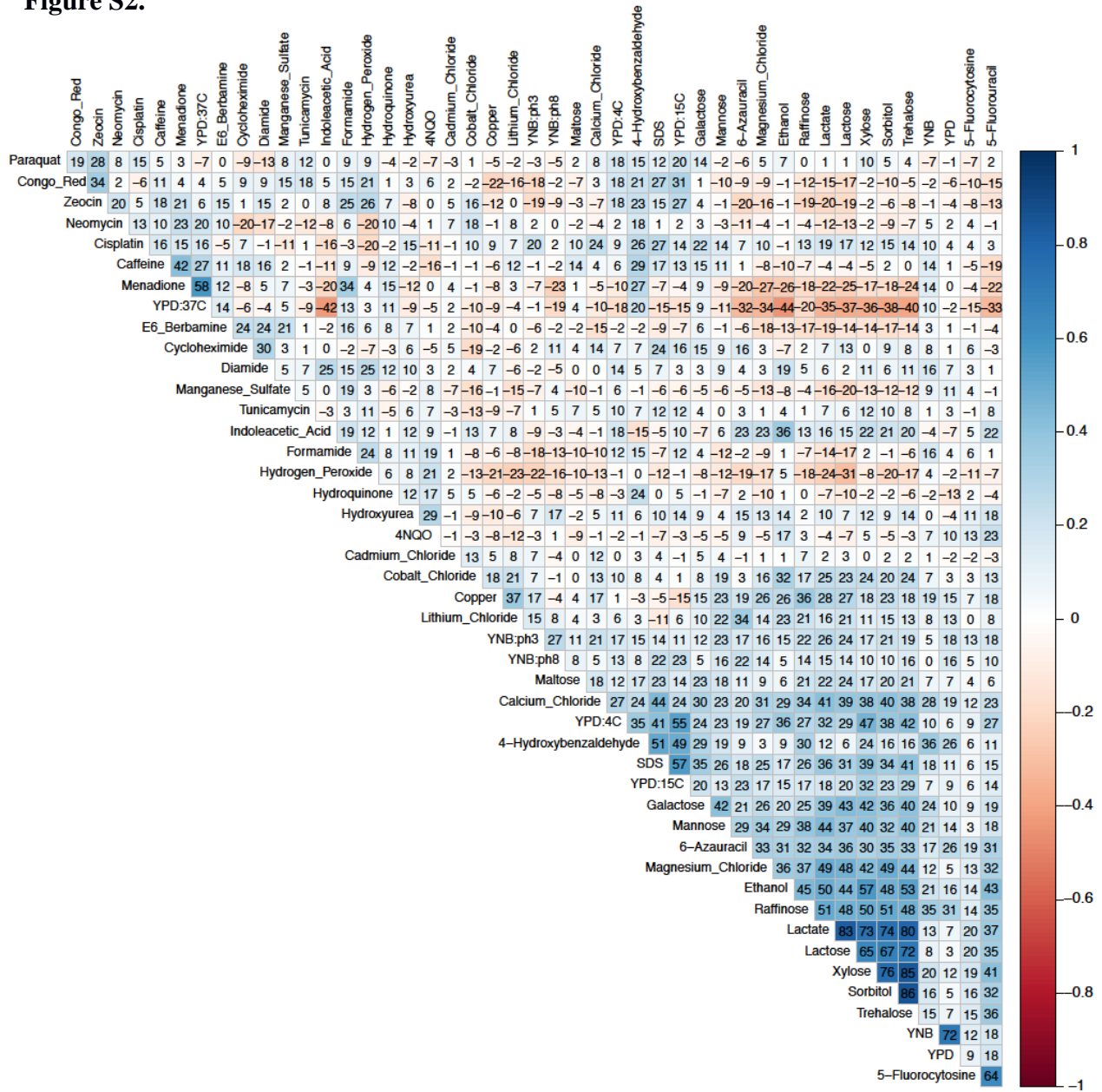

The Spearman correlation coefficients for all pairs of traits are shown. Numbers in table cells indicate (100 * correlation coefficient).

**Figure S3.**

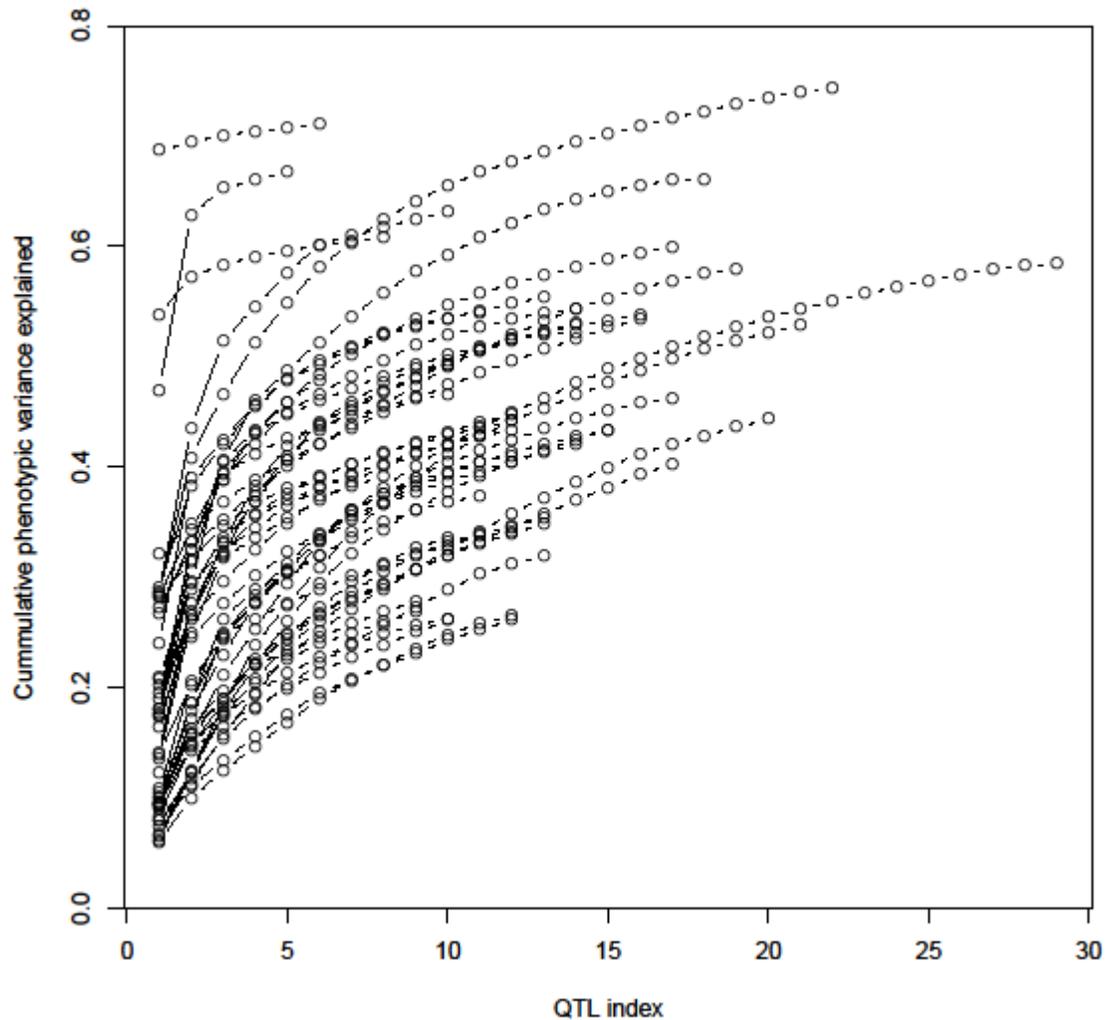

For each trait, the effect sizes of detected QTL are sorted from largest to smallest, and the cumulative phenotypic variance explained is plotted (Y-axis) against the number of detected QTL (X-axis). Under Orr's model, the curves should extrapolate to the origin; deviations from this expectation imply the presence of large-effect QTL for the corresponding traits.

**Supplementary Tables:**

**Table S1.** Drug doses, heritability statistics, and QTL summary statistics for traits investigated in this study.

**Table S2.** Table of detected QTL. Positions, effect sizes, confidence intervals and genes underneath detected QTL for each trait are listed.

**Author Contributions:**

Experiments were designed by J.S.B., I.M.E., and L.K. Experiments were performed by J.S.B., I.M.E., W.L., and T.L. Analyses were conducted by J.S.B. The manuscript was written by J.S.B. and L.K. and incorporates comments by all other authors.